\documentclass[prc,twocolumn,amsmath,amssymb,showpacs]{revtex4-1}

\usepackage{graphicx}
\usepackage{dcolumn}
\usepackage{bm}
\usepackage{epsfig}

\begin{document}

\preprint{draft}

\title{Suppression of Excited-State Contributions to Stellar Reaction Rates}

\author{T.\,Rauscher}%
\email{t.rauscher@herts.ac.uk}
\affiliation{%
Centre for Astrophysics Research, School of Physics, Astronomy and Mathematics, University of Hertfordshire, Hatfield AL10 9AB, UK}
\affiliation{%
Institute of Nuclear Research (ATOMKI), H-4001 Debrecen, POB.51., Hungary}%
\affiliation{%
Department of Physics, University of Basel, CH-4056 Basel, Switzerland}%

\date{\today}

\begin{abstract}
It has been shown in previous work [Phys.\ Rev.\ Lett.\ {\bf 101}, 191101 (2008); Phys.\ Rev.\ C {\bf 80}, 035801 (2009)] that a suppression of the stellar enhancement factor (SEF) occurs in some
endothermic reactions at and far from stability. This effect is re-evaluated using the ground-state contributions to the stellar reaction rates, which were shown to be better suited to judge the importance of excited state contributions than the previously applied SEFs. An update of the tables shown in Phys.\ Rev.\ C {\bf 80}, 035801 (2009) is given. The new evalution finds 2350 cases (out of a full set of 57513 reactions) for which the ground-state contribution is larger in the reaction direction with negative reaction $Q$ value than in the exothermic direction, thus providing exceptions to the commonly applied $Q$ value rule. The results confirm the Coulomb suppression effect but lead to a larger number of exceptions than previously found. This is due to the fact that often a large variation in the g.s.\ contribution does not lead to a sizeable change in the SEF. On the other hand, several previously identified cases do not appear anymore because it is found that their g.s.\ contribution is smaller than inferred from the SEF.
\end{abstract}


\maketitle

\section{Introduction}
\label{sec:intro}

Astrophysical reaction rates are central to any investigation in nucleosynthesis as they provide the quantitative description of temporal abundance changes in a hot plasma. They are related to reaction cross sections which, in turn, may be predicted in theoretical models or extracted from experiments. Compared to standard nuclear physics investigations of reactions on the ground state (g.s.) of nuclei, a treatment of reaction rates is complicated by the thermal population of excited states in the nuclei which are in thermal equilibrium with their hot environment. On the other hand, as long as thermal equilibrium is upheld, there is a mathematically rigorous reciprocity relation connecting the rates of forward and reverse reaction \cite{fow74,hol76,raureview}. This implies that only the \textit{stellar rate} for one direction, say, $A(a,b)B$ has to be determined by theoretical or experimental means. The \textit{stellar rate} for its inverse reaction $B(b,a)A$ automatically follows by applying the simple reciprocity relations.

Determining contributions of reactions proceeding on excited states of nuclei to the stellar rate remains an experimental challenge. Therefore it is interesting to theoretically investigate not only the magnitude of the contributions for each reaction but also which reaction direction is impacted less by such contributions. A determination of the rate in that direction would then be closer to the actually required stellar value. It is easy to show that fewer excited state transitions are involved in a stellar rate of a reaction with positive reaction $Q$ value than in one with negative $Q$ value. This is especially important for intermediate mass and heavy nuclei, whereas in light nuclei excited state ontributions are often small in either reaction direction, anway. The well-known ``$Q$ value rule'' follows directly, stating that for astrophysical purposes it is preferrable to use the direction of positive $Q$ value \cite{hol76,raureview,coulsupplett,coulsupp}. It has been shown recently, however, that an effect termed ``Coulomb suppression of excited state contributions'' can lead to a violation of this rule when the Coulomb barriers in entrance and exit channels are very different \cite{coulsupplett,coulsupp}. Since the relevant contributions of reactions on excited states involve different relative interaction energies, depending on the excitation energies of the participating excited states and the reaction $Q$ value, Coulomb barriers in entrance and exit channel act differently.

Ref.\ \cite{coulsupp} provided an in-depth investigation of this effect across the nuclear chart and also discussed the relevance in nucleosynthesis studies. It focused, however, on a comparison of the stellar enhancement factors (SEFs) in the forward and reverse reaction direction. Later it was shown that the SEF is not always a good measure of how reactions on excited states contribute to the stellar rate \cite{xfactor}. In this work, a similar investigation of the suppression effect is performed but using the better suited g.s.\  and excited state contributions to the stellar rate to check whether the conclusions of the previous investigation still hold and to provide updated tables.

\section{Definitions}
\label{sec:defs}

The SEF $f$ is defined as the ratio of the stellar rate $r^*$ and reactivity $R^*$ relative to the ground state (g.s.) rate $r^\mathrm{g.s.}$ and reactivity $R_0$, respectively, \cite{raureview,baoetal}
\begin{equation}
\label{eq:sef}
f(T)=\frac{r^*(T)}{r^\mathrm{g.s.}(T)}=\frac{R^*(T)}{R_0(T)}
\end{equation}
at a plasma temperature $T$.
While Coulomb suppression certainly drives the SEF value towards unity, a value close to 1.0 cannot be interpreted \textit{a priori}  as that excited state contributions are negligible. Later it has been shown that the more useful quantity to consider is the g.s.\ contribution to the stellar rate, \cite{xfactor,xfactlett}
\begin{equation}
\label{eq:xfactor}
X_0(T)=\left(2J_0+1\right) \frac{R_0}{G(T)R^*} \quad,
\end{equation}
where $J_0$ is the g.s.\ spin of the target nucleus and $G$ is the temperature-dependent partition function, defined as
\begin{equation}
G(T)=\sum_i (2J_i+1)e^{-\frac{E_i}{kT}} \quad.
\end{equation}
The sum runs over g.s.\ ($i=0$) and excited states ($i>0$) at excitation energy $E_i$.

The SEF can be close to unity even when excited state contributions are sizeable whereas $X_0$ directly shows the impact of such contributions, being $X_0 \approx 1$ when they are negligible and $X_0 \ll 1$ when they dominate the rate \cite{xfactor}.
Laboratory measurements only provide cross sections for nuclei in their g.s., and thus $r^\mathrm{g.s.}$, except for $^{180m}$Ta which naturally occurs in its long-lived isomer.
Therefore, $X_0$ is the quantity of interest also to experimentalists because it shows what fraction of the stellar rate can be constrained by measurements.

\section{Suppression of $X_0$}
\label{sec:xsupp}

\begin{figure}
\resizebox{\columnwidth}{!}{\rotatebox{270}{\includegraphics[clip=]{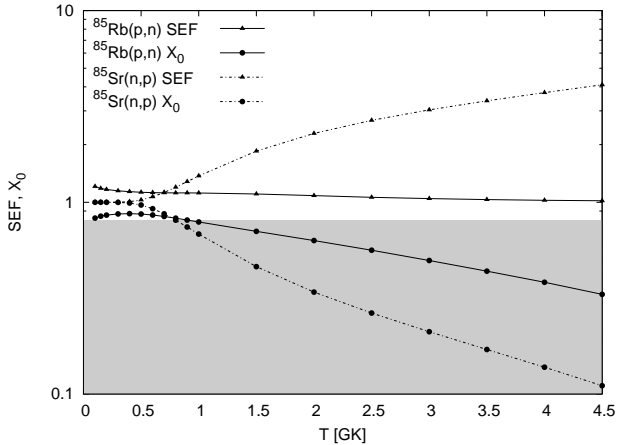}}}
\caption{Ground-state contributions $X_0$ and SEFs $f$ as function of plasma temperature $T$ for $^{85}$Sr(n,p)$^{85}$Rb and its reverse reaction. The exclusion limit $X_0<0.8$ is marked by the shaded area.\label{fig:rb85}}
\end{figure}

\begin{figure}
\resizebox{\columnwidth}{!}{\rotatebox{270}{\includegraphics[clip=]{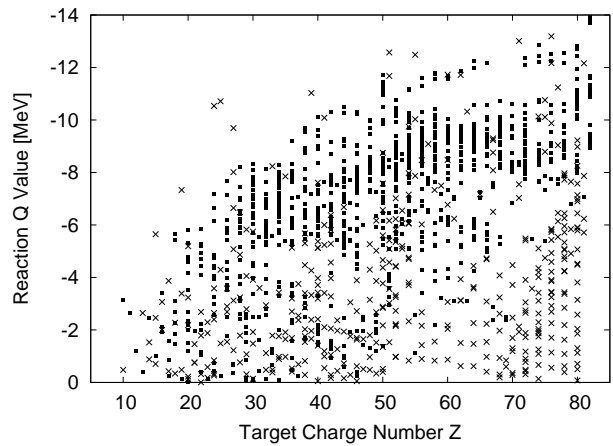}}}
\caption{Reaction $Q$ values for ($\alpha$,n) reactions (crosses) and ($\alpha$,p) reactions (squares) with $X_0^\mathrm{rev}>X_0^\mathrm{forw}$. To be compared to Fig.\ 2 of \cite{coulsupp}.\label{fig:qdep}}
\end{figure}

\begingroup
\squeezetable
\begin{table*}
\caption{\label{tab:ag} Targets for ($\alpha$,$\gamma$) reactions with negative $Q$ value but larger $X_0$ than their reverse reaction. Stable or long-lived targets are in italics. Underlined targets were also found in \cite{coulsupp}.}
\begin{ruledtabular}
\begin{tabular}{cccccccccccccccc}
$^{80}$Mo & \underline{$^{106}$Sb} & $^{116}$Xe & \underline{$^{123}$Ba} & \underline{$^{131}$Ce} & \underline{$^{136}$Nd} & \underline{$^{143}$Sm} & \underline{$^{147}$Gd} & \underline{$^{146}$Dy} & $^{169}$Ho & \underline{$^{154}$Yb} & \textit{$^\textit{179}$Hf} & \textit{$^\textit{185}$Re} & \textit{$^\textit{191}$Ir} & $^{193}$Au & $^{210}$Hg \\
\underline{$^{98}$Cd} & \underline{$^{107}$Sb} & $^{117}$Xe & \underline{$^{124}$Ba} & \underline{$^{132}$Ce} & \underline{$^{137}$Nd} & \underline{\textit{$^\textit{144}$Sm}} & \underline{$^{148}$Gd} & \underline{$^{148}$Dy} & $^{142}$Er & \underline{$^{156}$Yb} & \textit{$^\textit{180}$Hf} & \textit{$^\textit{187}$Re} & $^{192}$Ir & $^{194}$Au & $^{211}$Hg \\
$^{99}$Cd & \underline{$^{108}$Sb} & \underline{$^{118}$Xe} & \underline{$^{125}$Ba} & \underline{$^{133}$Ce} & \underline{$^{138}$Nd} & \underline{$^{145}$Sm} & \underline{$^{149}$Gd} & \underline{$^{149}$Dy} & $^{144}$Er & \underline{$^{158}$Yb} & $^{181}$Hf & $^{188}$Re & \textit{$^\textit{193}$Ir} & $^{195}$Au & $^{212}$Hg \\
$^{100}$Cd & \underline{$^{109}$Sb} & \underline{$^{119}$Xe} & \underline{$^{126}$Ba} & \underline{$^{134}$Ce} & \underline{$^{139}$Nd} & $^{146}$Sm & $^{150}$Gd & \underline{$^{150}$Dy} & $^{146}$Er & \underline{$^{160}$Yb} & $^{182}$Hf & $^{189}$Re & $^{194}$Ir & \textit{$^\textit{197}$Au} & $^{213}$Hg \\
$^{101}$Cd & \underline{$^{111}$Sb} & \underline{$^{120}$Xe} & \underline{$^{127}$Ba} & \underline{$^{135}$Ce} & $^{141}$Nd & \textit{$^\textit{147}$Sm} & $^{151}$Gd & \underline{$^{151}$Dy} & $^{148}$Er & \underline{$^{162}$Yb} & $^{183}$Hf & $^{190}$Re & $^{195}$Ir & $^{198}$Au & $^{214}$Hg \\
\underline{$^{99}$In} & \underline{$^{112}$Sb} & \underline{$^{121}$Xe} & \underline{$^{128}$Ba} & $^{139}$Ce & \underline{\textit{$^\textit{142}$Nd}} & \textit{$^\textit{148}$Sm} & \textit{$^\textit{152}$Gd} & \underline{$^{152}$Dy} & \underline{$^{150}$Er} & \textit{$^\textit{173}$Yb} & $^{184}$Hf & $^{201}$Re & $^{203}$Ir & $^{199}$Au & $^{215}$Hg \\
\underline{$^{101}$In} & \underline{$^{113}$Sb} & \underline{$^{122}$Xe} & \underline{\textit{$^\textit{130}$Ba}} & \underline{\textit{$^\textit{140}$Ce}} & \textit{$^\textit{143}$Nd} & \textit{$^\textit{150}$Sm} & \underline{$^{153}$Gd} & \underline{$^{154}$Dy} & \underline{$^{151}$Er} & \textit{$^\textit{174}$Yb} & \underline{$^{155}$Ta} & $^{167}$Os & $^{205}$Ir & $^{201}$Au & $^{216}$Hg \\
\underline{$^{102}$In} & $^{114}$Sb & \underline{$^{123}$Xe} & \underline{\textit{$^\textit{138}$Ba}} & $^{141}$Ce & \textit{$^\textit{144}$Nd} & \underline{$^{137}$Eu} & \textit{$^\textit{154}$Gd} & \underline{\textit{$^\textit{156}$Dy}} & \underline{$^{152}$Er} & $^{175}$Yb & $^{160}$Ta & $^{168}$Os & $^{207}$Ir & $^{205}$Au & $^{218}$Hg \\
\underline{$^{103}$In} & \underline{$^{115}$Sb} & \textit{$^\textit{136}$Xe} & $^{139}$Ba & \textit{$^\textit{142}$Ce} & \textit{$^\textit{145}$Nd} & $^{139}$Eu & \underline{\textit{$^\textit{155}$Gd}} & \underline{$^{157}$Dy} & \underline{$^{153}$Er} & \textit{$^\textit{176}$Yb} & $^{179}$Ta & $^{169}$Os & \underline{$^{168}$Pt} & $^{207}$Au & $^{220}$Hg \\
$^{104}$In & $^{116}$Sb & $^{137}$Xe & $^{140}$Ba & $^{143}$Ce & \textit{$^\textit{146}$Nd} & $^{141}$Eu & \textit{$^\textit{156}$Gd} & \textit{$^\textit{158}$Dy} & \underline{$^{154}$Er} & $^{177}$Yb & $^{183}$Ta & \textit{$^\textit{188}$Os} & \underline{$^{170}$Pt} & $^{209}$Au & $^{177}$Tl \\
\underline{$^{105}$In} & \underline{$^{133}$Sb} & $^{111}$Cs & $^{141}$Ba & $^{144}$Ce & $^{147}$Nd & $^{142}$Eu & \textit{$^\textit{157}$Gd} & \underline{$^{159}$Dy} & \underline{$^{156}$Er} & \underline{$^{178}$Yb} & $^{184}$Ta & \textit{$^\textit{189}$Os} & $^{188}$Pt & $^{213}$Au & $^{179}$Tl \\
$^{106}$In & $^{103}$Te & $^{117}$Cs & $^{127}$La & \underline{$^{129}$Pr} & \textit{$^\textit{148}$Nd} & \underline{$^{143}$Eu} & \textit{$^\textit{158}$Gd} & \textit{$^\textit{160}$Dy} & \underline{$^{157}$Er} & $^{180}$Yb & $^{185}$Ta & \textit{$^\textit{190}$Os} & \textit{$^\textit{190}$Pt} & $^{215}$Au & $^{181}$Tl \\
\underline{$^{107}$In} & \underline{$^{104}$Te} & $^{119}$Cs & $^{128}$La & \underline{$^{131}$Pr} & \underline{$^{131}$Pm} & $^{144}$Eu & $^{142}$Tb & \textit{$^\textit{161}$Dy} & \underline{$^{158}$Er} & $^{182}$Yb & $^{157}$W & $^{191}$Os & $^{191}$Pt & $^{174}$Hg & $^{190}$Tl \\
$^{108}$In & \underline{$^{106}$Te} & $^{120}$Cs & \underline{$^{129}$La} & $^{132}$Pr & \underline{$^{133}$Pm} & \underline{$^{145}$Eu} & $^{144}$Tb & \textit{$^\textit{162}$Dy} & \textit{$^\textit{164}$Er} & $^{184}$Yb & \underline{$^{158}$W } & \textit{$^\textit{192}$Os} & \textit{$^\textit{192}$Pt} & $^{177}$Hg & $^{195}$Tl \\
\underline{$^{99}$Sn} & $^{108}$Te & \underline{$^{121}$Cs} & $^{130}$La & \underline{$^{133}$Pr} & \underline{$^{137}$Pm} & \underline{$^{147}$Eu} & $^{146}$Tb & \textit{$^\textit{163}$Dy} & $^{165}$Er & $^{151}$Lu & $^{159}$W & $^{193}$Os & \textit{$^\textit{194}$Pt} & $^{178}$Hg & $^{196}$Tl \\
\underline{$^{102}$Sn} & \underline{$^{110}$Te} & $^{122}$Cs & \underline{$^{131}$La} & \underline{$^{135}$Pr} & $^{139}$Pm & $^{148}$Eu & $^{150}$Tb & \textit{$^\textit{164}$Dy} & \textit{$^\textit{166}$Er} & \underline{$^{153}$Lu} & \underline{$^{160}$W } & $^{194}$Os & \textit{$^\textit{195}$Pt} & \underline{$^{188}$Hg} & $^{197}$Tl \\
\underline{$^{104}$Sn} & \underline{$^{111}$Te} & \underline{$^{123}$Cs} & \textit{$^\textit{138}$La} & $^{136}$Pr & $^{140}$Pm & \underline{$^{149}$Eu} & $^{152}$Tb & $^{165}$Dy & \textit{$^\textit{167}$Er} & \underline{$^{155}$Lu} & $^{163}$W & $^{201}$Os & \textit{$^\textit{196}$Pt} & \underline{$^{190}$Hg} & $^{198}$Tl \\
\underline{$^{105}$Sn} & \underline{$^{112}$Te} & $^{124}$Cs & \textit{$^\textit{139}$La} & $^{137}$Pr & $^{141}$Pm & $^{150}$Eu & $^{153}$Tb & $^{166}$Dy & \textit{$^\textit{168}$Er} & \underline{$^{167}$Lu} & $^{165}$W & $^{202}$Os & $^{197}$Pt & \underline{$^{192}$Hg} & $^{199}$Tl \\
\underline{$^{106}$Sn} & \underline{$^{114}$Te} & \underline{$^{125}$Cs} & $^{140}$La & $^{140}$Pr & $^{142}$Pm & \textit{$^\textit{151}$Eu} & $^{155}$Tb & $^{145}$Ho & $^{169}$Er & \textit{$^\textit{175}$Lu} & $^{181}$W & $^{203}$Os & \textit{$^\textit{198}$Pt} & \underline{$^{194}$Hg} & $^{200}$Tl \\
\underline{$^{107}$Sn} & \underline{$^{115}$Te} & $^{126}$Cs & $^{141}$La & \underline{\textit{$^\textit{141}$Pr}} & \underline{$^{143}$Pm} & $^{152}$Eu & $^{156}$Tb & $^{147}$Ho & \textit{$^\textit{170}$Er} & \textit{$^\textit{176}$Lu} & \textit{$^\textit{184}$W } & $^{204}$Os & $^{203}$Pt & \textit{$^\textit{196}$Hg} & $^{201}$Tl \\
\underline{$^{108}$Sn} & \underline{$^{116}$Te} & \underline{$^{127}$Cs} & $^{142}$La & $^{143}$Pr & $^{147}$Pm & \textit{$^\textit{153}$Eu} & $^{157}$Tb & \underline{$^{150}$Ho} & $^{171}$Er & $^{177}$Lu & $^{185}$W & $^{205}$Os & $^{204}$Pt & \textit{$^\textit{198}$Hg} & \textit{$^\textit{203}$Tl} \\
\underline{$^{110}$Sn} & \underline{$^{117}$Te} & $^{137}$Cs & $^{143}$La & $^{145}$Pr & $^{148}$Pm & $^{132}$Gd & $^{158}$Tb & \underline{$^{151}$Ho} & $^{172}$Er & $^{178}$Lu & \textit{$^\textit{186}$W } & $^{206}$Os & $^{205}$Pt & \underline{\textit{$^\textit{199}$Hg}} & $^{204}$Tl \\
\underline{$^{111}$Sn} & \underline{$^{117}$I } & $^{138}$Cs & $^{116}$Ce & $^{146}$Pr & $^{149}$Pm & $^{136}$Gd & \textit{$^\textit{159}$Tb} & $^{153}$Ho & $^{145}$Tm & $^{179}$Lu & $^{187}$W & $^{207}$Os & $^{206}$Pt & \textit{$^\textit{200}$Hg} & \textit{$^\textit{205}$Tl} \\
\underline{\textit{$^\textit{112}$Sn}} & \underline{$^{119}$I } & $^{139}$Cs & $^{118}$Ce & $^{122}$Nd & $^{126}$Sm & $^{138}$Gd & $^{160}$Tb & \underline{$^{159}$Ho} & $^{146}$Tm & $^{152}$Hf & $^{188}$W & $^{208}$Os & $^{207}$Pt & \textit{$^\textit{202}$Hg} & $^{213}$Tl \\
\underline{$^{113}$Sn} & $^{135}$I & $^{112}$Ba & $^{120}$Ce & $^{124}$Nd & $^{128}$Sm & $^{139}$Gd & $^{161}$Tb & $^{161}$Ho & $^{147}$Tm & \underline{$^{154}$Hf} & $^{200}$W & $^{168}$Ir & $^{208}$Pt & \underline{$^{203}$Hg} & $^{215}$Tl \\
\underline{\textit{$^\textit{114}$Sn}} & $^{108}$Xe & $^{114}$Ba & $^{124}$Ce & $^{126}$Nd & \underline{$^{136}$Sm} & $^{140}$Gd & $^{162}$Tb & $^{162}$Ho & $^{152}$Tm & $^{155}$Hf & $^{201}$W & $^{171}$Ir & $^{209}$Pt & \textit{$^\textit{204}$Hg} & $^{219}$Tl \\
\underline{\textit{$^\textit{115}$Sn}} & $^{110}$Xe & $^{116}$Ba & $^{125}$Ce & $^{131}$Nd & \underline{$^{137}$Sm} & $^{141}$Gd & $^{136}$Dy & $^{163}$Ho & $^{170}$Tm & \underline{$^{156}$Hf} & $^{202}$W & \underline{$^{172}$Ir} & $^{210}$Pt & $^{205}$Hg &  \\
$^{101}$Sb & $^{112}$Xe & $^{118}$Ba & \underline{$^{126}$Ce} & \underline{$^{132}$Nd} & \underline{$^{138}$Sm} & \underline{$^{142}$Gd} & $^{138}$Dy & $^{164}$Ho & $^{172}$Tm & \underline{$^{157}$Hf} & $^{203}$W & $^{173}$Ir & $^{211}$Pt & $^{206}$Hg &  \\
$^{102}$Sb & $^{113}$Xe & $^{120}$Ba & \underline{$^{128}$Ce} & \underline{$^{133}$Nd} & \underline{$^{139}$Sm} & \underline{$^{143}$Gd} & $^{140}$Dy & \textit{$^\textit{165}$Ho} & $^{148}$Yb & \underline{$^{158}$Hf} & $^{164}$Re & \underline{$^{175}$Ir} & $^{212}$Pt & $^{207}$Hg &  \\
$^{103}$Sb & $^{114}$Xe & $^{121}$Ba & \underline{$^{129}$Ce} & \underline{$^{134}$Nd} & \underline{$^{140}$Sm} & \underline{$^{144}$Gd} & \underline{$^{142}$Dy} & $^{167}$Ho & \underline{$^{152}$Yb} & \textit{$^\textit{177}$Hf} & \underline{$^{167}$Re} & \underline{$^{177}$Ir} & $^{214}$Pt & $^{208}$Hg &  \\
\underline{$^{105}$Sb} & $^{115}$Xe & $^{122}$Ba & \underline{$^{130}$Ce} & \underline{$^{135}$Nd} & \underline{$^{142}$Sm} & \underline{$^{146}$Gd} & \underline{$^{144}$Dy} & $^{168}$Ho & \underline{$^{153}$Yb} & \textit{$^\textit{178}$Hf} & $^{182}$Re & $^{189}$Ir & $^{172}$Au & $^{209}$Hg &
\end{tabular}
\end{ruledtabular}
\end{table*}
\endgroup

\begingroup
\squeezetable
\begin{table}
\caption{\label{tab:pg} Same as Table \ref{tab:ag} but for (p,$\gamma$).}
\begin{ruledtabular}
\begin{tabular}{cccccccc}
\underline{$^{33}$Ar} & \underline{$^{62}$Ge} & $^{94}$Cd & \underline{$^{113}$Xe} & $^{138}$Gd & $^{152}$Hf & $^{168}$Os & $^{179}$Pb \\
\underline{$^{37}$Ca} & $^{63}$Ge & \underline{$^{95}$Cd} & \underline{$^{114}$Xe} & \underline{$^{138}$Dy} & \underline{$^{154}$Hf} & \underline{$^{170}$Os} & $^{180}$Pb \\
\underline{$^{40}$Ti} & $^{64}$Ge & \underline{$^{96}$Cd} & $^{111}$Cs & $^{140}$Dy & $^{155}$Hf & $^{168}$Pt & $^{181}$Pb \\
$^{41}$Ti & $^{68}$Se & $^{94}$In & $^{112}$Cs & $^{142}$Dy & $^{156}$Hf & $^{170}$Pt & $^{182}$Pb \\
$^{45}$Fe & $^{72}$Kr & $^{98}$Sn & $^{112}$Ba & \underline{$^{144}$Dy} & \underline{$^{157}$Hf} & $^{172}$Pt & $^{183}$Pb \\
\underline{$^{48}$Fe} & $^{74}$Sr & \underline{$^{100}$Sn} & $^{116}$Ba & $^{142}$Er & $^{158}$Hf & $^{174}$Pt & $^{184}$Pb \\
\underline{$^{49}$Fe} & $^{75}$Sr & $^{101}$Sn & $^{118}$Ba & $^{146}$Er & $^{156}$W & \underline{$^{176}$Pt} & $^{185}$Pb \\
\underline{$^{52}$Ni} & \underline{$^{76}$Sr} & \underline{$^{102}$Sn} & $^{118}$Ce & \underline{$^{148}$Er} & $^{157}$W & $^{173}$Hg & \underline{$^{186}$Pb} \\
\underline{$^{53}$Ni} & $^{73}$Y & \underline{$^{103}$Sn} & $^{122}$Ce & $^{145}$Tm & \underline{$^{158}$W } & $^{174}$Hg & $^{187}$Pb \\
\underline{$^{54}$Ni} & $^{81}$Nb & \underline{$^{104}$Te} & $^{124}$Nd & $^{146}$Tm & $^{159}$W & $^{176}$Hg & \underline{$^{188}$Pb} \\
$^{54}$Zn & $^{84}$Mo & \underline{$^{106}$Te} & \underline{$^{126}$Nd} & $^{148}$Yb & \underline{$^{160}$W } & $^{177}$Hg &  \\
\underline{$^{56}$Zn} & $^{83}$Tc & $^{107}$Te & $^{123}$Pm & \underline{$^{150}$Yb} & $^{162}$W & $^{178}$Hg &  \\
\underline{$^{57}$Zn} & $^{84}$Tc & \underline{$^{108}$Te} & $^{132}$Sm & \underline{$^{151}$Yb} & $^{164}$W & \underline{$^{180}$Hg} &  \\
\underline{$^{58}$Zn} & \underline{$^{86}$Ru} & \underline{$^{109}$Te} & $^{134}$Sm & \underline{$^{152}$Yb} & $^{162}$Os & \underline{$^{181}$Hg} &  \\
\underline{$^{60}$Ge} & \underline{$^{90}$Pd} & $^{106}$I & $^{134}$Gd & \underline{$^{153}$Yb} & \underline{$^{164}$Os} & $^{175}$Tl &  \\
\underline{$^{61}$Ge} & \underline{$^{92}$Pd} & \underline{$^{112}$Xe} & \underline{$^{136}$Gd} & $^{154}$Yb & $^{166}$Os & $^{178}$Pb &
\end{tabular}
\end{ruledtabular}
\end{table}
\endgroup

Here, an investigation of the suppression effect is performed similar to the one shown in \cite{coulsupp} but focussing on $X_0$, comparing the g.s.\ contribution of the forward rate $X_0^\mathrm{forw}$ to that for its reverse rate $X_0^\mathrm{rev}$ and identifying cases with $X_0^\mathrm{rev}>X_0^\mathrm{forw}$. The calculations were performed with the Hauser-Feshbach code SMARAGD \cite{smaragd} using updated excited state information \cite{nudat,ensdf} and masses \cite{cyburt}. The code also uses an improved barrier penetration routine which is appropriate to treat low-energy $\alpha$ transmissions through high Coulomb barriers \cite{raureview}. As in the previous work, the forward reaction direction is defined as the one with positive reaction $Q$ value, and reactions involving light projectiles (nucleons, $\alpha$) are studied, with target nuclei ranging from Ne to Bi between the proton and neutron driplines. The selection criteria for the displayed cases had to be modified with respect to the ones used in \cite{coulsupp} because of the different properties of $X_0$. To avoid trivial cases and find sizeable differences, only reactions with $X_0^\mathrm{rev}/X_0^\mathrm{forw}>1.3$ and $X_0^\mathrm{rev}>0.8$ were considered. This choice ensures that the stellar rate of the reaction with negative $Q$ value is dominated by the g.s.\ contribution. It also implies that the g.s.\ contributions differ by at least 30\% between forward and reverse direction. Only plasma temperatures $T\leq 4.5$ GK were included to identify cases important in most nucleosynthesis environments and to eliminate those only occurring at very high temperature. 

Using these restrictions, 2350 reactions (out of 57513) exhibiting a strong suppression effect of $X_0^\mathrm{rev}$ remained, more than in the previous investigation. The reason for a larger number of cases being found is that a large variation in the g.s.\ contribution does not cause an equally big change in the SEF in many cases. On the other hand, some cases do not appear anymore because the g.s.\ contribution decreases although the SEF is remaining constant or becoming smaller. Figure \ref{fig:rb85} shows an example of this and also how the selection criteria work for the reaction $^{85}$Sr(n,p)$^{85}$Rb and its reverse reaction. Although the SEF of $^{85}$Rb(p,n)$^{85}$Sr ($Q=-1.847$ MeV) is close to unity for all temperatures, the actual g.s.\ contribution decreases with increasing temperature. Moreover, comparing the evolution of $X_0$ as function of temperature for forward and reverse reaction, it becomes apparent that $X_0^\mathrm{rev}$ is larger than $1.3X_0^\mathrm{forw}$ only when both are smaller than 0.8. At lower $T$, both g.s.\ contributions are above the cutoff at 0.8 but there $X_0^\mathrm{forw}>X_0^\mathrm{rev}$. Therefore, the reaction $^{85}$Rb(p,n)$^{85}$Sr does not show up anymore in the tables given below.

Figure \ref{fig:qdep} can be directly compared to Fig.\ 2 of \cite{coulsupp}, illustrating the dependence of the suppression on the Coulomb barrier. Shown is the
obtained range of $Q$ values for ($\alpha$,n) and ($\alpha$,p) reactions with negative $Q$ values fullfilling the above criteria. As explained in \cite{coulsupp},
larger maximal $\left| Q\right|$ is allowed with increasing charge $Z$. This documents that the explanation given in the earlier paper \cite{coulsupp} can also be applied when using g.s.\ contributions instead of SEF.

\section{Results and Discussion}

\begingroup
\squeezetable
\begin{table}
\caption{\label{tab:gn} Same as Table \ref{tab:ag} but for ($\gamma$,n).}
\begin{ruledtabular}
\begin{tabular}{cccccccc}
$^{29}$Ne & $^{44}$Mg & $^{61}$Ca & $^{105}$Se & \underline{$^{122}$Sr} & $^{169}$Te & $^{230}$Yb & $^{243}$Au \\
$^{31}$Ne & $^{22}$Al & $^{72}$Ti & $^{115}$Se & $^{124}$Zr & $^{173}$Ba & \underline{$^{236}$Hf} & $^{243}$Hg \\
$^{28}$Na & \underline{$^{40}$Al} & $^{77}$V & \underline{$^{116}$Se} & $^{128}$Mo & \underline{$^{186}$Ce} & $^{239}$Os & $^{245}$Hg \\
$^{35}$Mg & $^{45}$Al & $^{81}$Mn & $^{109}$Kr & $^{148}$Ru & \underline{$^{188}$Ce} & $^{244}$Os & $^{260}$Hg \\
$^{37}$Mg & $^{48}$Si & $^{101}$Ge & $^{118}$Kr & $^{149}$Rh & \underline{$^{188}$Nd} & $^{251}$Pt & $^{264}$Hg \\
\underline{$^{42}$Mg} & $^{55}$Ar & $^{108}$Ge & \underline{$^{121}$Rb} & \underline{$^{151}$Rh} & $^{190}$Sm & $^{252}$Pt &
\end{tabular}
\end{ruledtabular}
\end{table}
\endgroup

\begingroup
\squeezetable
\begin{table}
\caption{\label{tab:gpgananp} Same as Table \ref{tab:ag} but for ($\gamma$,p), (n,$\alpha$), and (n,$\gamma$) reactions.}
\begin{ruledtabular}
\begin{tabular}{cccccccc}
\multicolumn{8}{l}{($\gamma$,p):} \\
$^{38}$Ti & $^{46}$Fe & $^{53}$Co & $^{64}$Se & $^{99}$Sn & $^{137}$Eu & $^{165}$Ta &  \\
$^{46}$Cr & $^{52}$Co & $^{55}$Zn & $^{79}$Sr & $^{131}$Pm & $^{150}$Tb & $^{169}$Ta & \\
\multicolumn{8}{l}{(n,$\alpha$):} \\
\textit{$^\textit{65}$Cu} &  &  &  &  &  &  & \\
\multicolumn{8}{l}{(n,$\gamma$):} \\
$^{38}$Mg & $^{79}$Cr & $^{130}$Ru & $^{133}$Ag & $^{191}$Ce & $^{194}$Gd & $^{246}$Pt &  \\
$^{44}$Mg & $^{80}$Fe & $^{132}$Ru & \underline{$^{136}$Cd} & $^{192}$Sm & $^{196}$Gd & $^{243}$Hg &  \\
$^{40}$Si & $^{82}$Ni & \underline{$^{132}$Pd} & $^{158}$Sn & $^{189}$Eu & $^{222}$Yb & $^{263}$Tl &  \\
$^{74}$Cr & $^{91}$Cu & $^{144}$Pd & $^{173}$I & $^{192}$Gd & $^{231}$Lu &  & 
\end{tabular}
\end{ruledtabular}
\end{table}
\endgroup

\begingroup
\squeezetable
\begin{table}
\caption{\label{tab:pn} Same as Table \ref{tab:ag} but for (p,n).}
\begin{ruledtabular}
\begin{tabular}{cccccccc}
\underline{$^{32}$Si} & \underline{$^{77}$As} & \underline{\textit{$^\textit{104}$Ru}} & \textit{$^\textit{126}$Te} & \textit{$^\textit{140}$Ce} & \textit{$^\textit{163}$Dy} & \textit{$^\textit{180}$Ta} & \underline{\textit{$^\textit{202}$Hg}} \\
$^{36}$Cl & $^{79}$Se & \underline{$^{106}$Ru} & $^{127}$Te & $^{141}$Ce & \underline{$^{166}$Dy} & $^{174}$W & \underline{\textit{$^\textit{204}$Hg}} \\
\textit{$^\textit{40}$Ar} & \underline{\textit{$^\textit{82}$Se}} & \underline{$^{105}$Rh} & \underline{\textit{$^\textit{128}$Te}} & \underline{\textit{$^\textit{142}$Ce}} & $^{162}$Ho & \underline{$^{188}$W } & $^{201}$Tl \\
\underline{$^{42}$Ar} & \underline{\textit{$^\textit{81}$Br}} & \underline{$^{107}$Pd} & \underline{\textit{$^\textit{130}$Te}} & \underline{$^{144}$Ce} & \underline{\textit{$^\textit{165}$Ho}} & $^{182}$Re & $^{202}$Tl \\
\textit{$^\textit{46}$Ca} & \underline{$^{85}$Kr} & \underline{\textit{$^\textit{110}$Pd}} & \underline{$^{132}$Te} & \textit{$^\textit{143}$Nd} & \textit{$^\textit{167}$Er} & $^{184}$Re & \underline{\textit{$^\textit{203}$Tl}} \\
\underline{\textit{$^\textit{48}$Ca}} & \underline{\textit{$^\textit{86}$Kr}} & \underline{$^{112}$Pd} & $^{124}$I & \underline{\textit{$^\textit{144}$Nd}} & $^{169}$Er & \underline{\textit{$^\textit{185}$Re}} & \underline{$^{204}$Tl} \\
\underline{$^{47}$Sc} & \underline{$^{87}$Rb} & \textit{$^\textit{113}$Cd} & \textit{$^\textit{131}$Xe} & \textit{$^\textit{145}$Nd} & $^{164}$Yb & \underline{\textit{$^\textit{187}$Re}} & \underline{\textit{$^\textit{205}$Tl}} \\
\underline{\textit{$^\textit{49}$Ti}} & \textit{$^\textit{87}$Sr} & \underline{\textit{$^\textit{116}$Cd}} & $^{133}$Xe & \underline{\textit{$^\textit{146}$Nd}} & $^{175}$Yb & $^{191}$Os & $^{200}$Pb \\
\textit{$^\textit{53}$Cr} & \textit{$^\textit{88}$Sr} & \underline{$^{118}$Cd} & \underline{\textit{$^\textit{134}$Xe}} & \underline{\textit{$^\textit{148}$Nd}} & \underline{\textit{$^\textit{176}$Yb}} & \underline{$^{194}$Os} & \textit{$^\textit{206}$Pb} \\
\underline{$^{60}$Fe} & \underline{$^{90}$Sr} & $^{111}$In & \underline{\textit{$^\textit{136}$Xe}} & \underline{\textit{$^\textit{150}$Nd}} & \underline{$^{178}$Yb} & $^{186}$Ir & \textit{$^\textit{207}$Pb} \\
\textit{$^\textit{59}$Co} & \textit{$^\textit{91}$Zr} & \underline{\textit{$^\textit{113}$In}} & $^{128}$Cs & \underline{$^{147}$Pm} & $^{163}$Lu & $^{188}$Ir & \textit{$^\textit{208}$Pb} \\
\underline{\textit{$^\textit{64}$Ni}} & \textit{$^\textit{92}$Zr} & \underline{\textit{$^\textit{115}$In}} & $^{130}$Cs & $^{145}$Sm & $^{174}$Lu & \underline{\textit{$^\textit{193}$Ir}} & $^{209}$Pb \\
\underline{$^{66}$Ni} & \underline{$^{93}$Zr} & \textit{$^\textit{118}$Sn} & \underline{$^{135}$Cs} & \underline{$^{156}$Sm} & \underline{\textit{$^\textit{175}$Lu}} & \textit{$^\textit{195}$Pt} & $^{210}$Pb \\
\underline{\textit{$^\textit{65}$Cu}} & \underline{\textit{$^\textit{94}$Zr}} & \underline{\textit{$^\textit{120}$Sn}} & \textit{$^\textit{135}$Ba} & $^{148}$Eu & \underline{$^{177}$Lu} & \underline{\textit{$^\textit{196}$Pt}} & $^{212}$Pb \\
\underline{$^{67}$Cu} & \underline{\textit{$^\textit{96}$Zr}} & \underline{\textit{$^\textit{122}$Sn}} & \textit{$^\textit{136}$Ba} & \underline{$^{155}$Eu} & $^{160}$Hf & \underline{$^{197}$Pt} &  \\
\underline{\textit{$^\textit{70}$Zn}} & \underline{\textit{$^\textit{97}$Mo}} & \underline{\textit{$^\textit{124}$Sn}} & \textit{$^\textit{137}$Ba} & $^{150}$Tb & \underline{\textit{$^\textit{179}$Hf}} & \underline{\textit{$^\textit{198}$Pt}} &  \\
\underline{$^{72}$Zn} & \textit{$^\textit{98}$Mo} & \underline{$^{126}$Sn} & \underline{\textit{$^\textit{138}$Ba}} & $^{152}$Tb & \underline{$^{182}$Hf} & \underline{$^{200}$Pt} &  \\
\underline{\textit{$^\textit{71}$Ga}} & \underline{\textit{$^\textit{100}$Mo}} & \underline{\textit{$^\textit{123}$Sb}} & \underline{\textit{$^\textit{139}$La}} & $^{156}$Tb & $^{167}$Ta & \underline{\textit{$^\textit{197}$Au}} &  \\
\underline{\textit{$^\textit{76}$Ge}} & \underline{$^{99}$Tc} & \underline{$^{125}$Sb} & $^{139}$Ce & \underline{$^{161}$Tb} & $^{174}$Ta & \underline{$^{199}$Au} &
\end{tabular}
\end{ruledtabular}
\end{table}
\endgroup

\begingroup
\squeezetable
\begin{table}
\caption{\label{tab:np} Same as Table \ref{tab:ag} but for (n,p) reactions.}
\begin{ruledtabular}
\begin{tabular}{cccccccc}
\textit{$^\textit{32}$S } & \textit{$^\textit{40}$Ca} &  &  &  &  &  &
\end{tabular}
\end{ruledtabular}
\end{table}
\endgroup

Tables \ref{tab:ag}--\ref{tab:pa} list the reactions found to have $X_0^\mathrm{rev}>X_0^\mathrm{forw}$ according to the criteria discussed above. Shown are the target nuclides for the endothermic direction. Stable or long-lived target nuclei are printed in italics. The present Tables \ref{tab:ag}$-$\ref{tab:ap} are in the same sequence as and formatted similarly to tables I--VIII in \cite{coulsupp} and supersede those. Underlined target nuclei appeared already in the previous tables. Most nuclei are new entries, they are not underlined. Target nuclei which do not appear anymore can be found by direct comparison with the previous tables. Noteable are the many more cases found for endothermic ($\alpha$,p) reactions (Table \ref{tab:ap}) compared to the previous investigation. Table \ref{tab:pa} is new because endothermic (p,$\alpha$) reactions did not show any SEF suppression previously but a suppression of the g.s.\ contribution was found here. A further difference to the previous investigation are ($\gamma$,$\alpha$) reactions on the following target nuclei: $^{27}$S, $^{29}$Ar, $^{38}$Sc, $^{48}$Mn, $^{83}$Tc, $^{29}$S, $^{36}$K, $^{40}$Sc, $^{57}$Ge, $^{97}$Sn. No SEF suppression was found for endothermic ($\gamma$,$\alpha$) but a suppression of $X_0$ was found for the above few cases.

The majority of the shown cases are caused by Coulomb suppression of the excited state contributions in one reaction channel. As in \cite{coulsupp}, however, a suppression is also found for a few (n,$\gamma$), ($\gamma$,n), ($\gamma$,p), ($\gamma$,$\alpha$), (n,p), (n,$\alpha$), and (p,$\alpha$) cases (Tables \ref{tab:gn}, \ref{tab:gpgananp}, \ref{tab:np}, and the ($\gamma$,$\alpha$) reactions mentioned above). Obviously, this suppression is not caused by the Coulomb barrier but rather transitions from excited states are suppressed by selection rules and centrifugal barriers. It has to be noted that the results for most of these nuclei -- as well as others far off stability -- are purely based on theory, as neither spectroscopic information nor masses are available from experiment. In these cases, recommended mass values were taken \cite{cyburt,ame} or a theoretical mass model \cite{frdm} was used to compute the separation energies. In the absence of knowledge about excited states, a theoretical nuclear level density was employed \cite{darko,rtk}. Therefore new mass measurements could change the reaction $Q$ values and also further experimental information on excited states may impact the results presented here, especially close to the driplines.

As already discussed in \cite{coulsupp}, to current knowledge most of the shown reactions are not of direct astrophysical importance. They are merely interesting cases to demonstrate the suppression effect and to study where the standard $Q$ value rule can be applied and where this is not possible. Future developments in astrophysical models and experimental techniques should not be precluded, however, and therefore the full tables are given. For further details on the astrophysical relevance and also on the limitations of the applied reaction model, see \cite{coulsupp}.

As has been pointed out previously, most remarkable in the current astrophysics context are the ($\alpha$,$\gamma$) reactions on proton-rich targets listed in Table \ref{tab:ag}, especially the ones on stable targets with neutron number $N\geq 82$. These are important to study ($\gamma$,$\alpha$) rates and the optical $\alpha$+nucleus potential required for predictions relevant to the synthesis of $p$ nuclei \cite{p-review}. Also important in the astrophysical $\gamma$- and in the $\nu$p process are (p,n) reactions with negative $Q$ value (Table \ref{tab:pn}) because (n,p) reactions on proton-rich isotopes play a role in these nucleosynthesis processes (see, e.g., \cite{coulsupplett,coulsupp,p-review,rapp,froh,rauomeg,frohomeg}). Endothermic ($\alpha$,p) reactions (Table \ref{tab:ap}) on proton-rich nuclei are also of interest for $\nu$p process studies \cite{almu}. Further of interest are the (p,$\gamma$) reactions along the proton dripline shown in Table \ref{tab:pg} because they appear in the rp-process \cite{somRPpaper}. Moreover, the large number of ($\alpha$,$\gamma$) and (p,$\gamma$) reactions (and the comparatively small number of ($\gamma$,p) and ($\gamma$,$\alpha$) in Table \ref{tab:gpgananp} and given in the text above) found in this study underlines the fact that it is almost always preferable to measure in the capture direction, even for endothermic captures. This is consistent with the findings of \cite{raureview,rausensi}.

\begingroup
\squeezetable
\begin{table*}
\caption{\label{tab:an} Same as Table \ref{tab:ag} but for ($\alpha$,n).}
\begin{ruledtabular}
\begin{tabular}{cccccccccccccccc}
\underline{\textit{$^\textit{22}$Ne}} & \underline{\textit{$^\textit{50}$Ti}} & \textit{$^\textit{62}$Ni} & \underline{$^{84}$Se} & \underline{$^{93}$Y } & \textit{$^\textit{97}$Mo} & \underline{$^{116}$Ag} & $^{113}$Sb & \underline{$^{137}$I } & \underline{$^{147}$Pr} & $^{174}$Dy & \underline{$^{196}$Yb} & $^{194}$W & \underline{$^{200}$Os} & \underline{$^{206}$Pt} & \underline{$^{212}$Hg} \\
\textit{$^\textit{27}$Al} & \underline{$^{52}$Ti} & \underline{\textit{$^\textit{64}$Ni}} & $^{75}$Br & \underline{$^{94}$Y } & \underline{\textit{$^\textit{98}$Mo}} & \underline{\textit{$^\textit{114}$Cd}} & $^{115}$Sb & \textit{$^\textit{126}$Xe} & $^{132}$Nd & $^{176}$Dy & \underline{$^{198}$Yb} & $^{196}$W & $^{201}$Os & $^{207}$Pt & \underline{$^{214}$Hg} \\
\textit{$^\textit{29}$Si} & $^{48}$V & $^{65}$Ni & \underline{$^{82}$Br} & \underline{$^{95}$Y } & \underline{\textit{$^\textit{100}$Mo}} & \underline{\textit{$^\textit{116}$Cd}} & \underline{$^{125}$Sb} & \textit{$^\textit{130}$Xe} & $^{138}$Nd & $^{178}$Dy & \underline{$^{200}$Yb} & \underline{$^{197}$W } & \underline{$^{202}$Os} & $^{208}$Pt & \underline{$^{215}$Hg} \\
\underline{$^{32}$Si} & \textit{$^\textit{50}$V } & \underline{$^{66}$Ni} & \underline{$^{83}$Br} & $^{96}$Y & \underline{$^{102}$Mo} & $^{117}$Cd & \underline{$^{127}$Sb} & \underline{\textit{$^\textit{134}$Xe}} & \textit{$^\textit{148}$Nd} & $^{180}$Dy & $^{167}$Lu & $^{198}$W & \underline{$^{203}$Os} & $^{209}$Pt & \underline{$^{216}$Hg} \\
\textit{$^\textit{31}$P } & \underline{\textit{$^\textit{51}$V }} & \underline{$^{68}$Ni} & \underline{$^{85}$Br} & \underline{$^{97}$Y } & \underline{$^{103}$Mo} & \underline{$^{118}$Cd} & \underline{$^{129}$Sb} & $^{135}$Xe & \underline{\textit{$^\textit{150}$Nd}} & $^{182}$Dy & \underline{$^{178}$Lu} & $^{199}$W & \underline{$^{204}$Os} & \underline{$^{210}$Pt} & $^{218}$Hg \\
$^{33}$P & \underline{$^{53}$V } & \textit{$^\textit{65}$Cu} & \underline{\textit{$^\textit{82}$Kr}} & $^{89}$Zr & \underline{$^{104}$Mo} & \underline{$^{120}$Cd} & \underline{$^{131}$Sb} & \underline{\textit{$^\textit{136}$Xe}} & $^{158}$Nd & \underline{$^{184}$Dy} & \underline{$^{185}$Lu} & \underline{$^{200}$W } & $^{205}$Os & $^{211}$Pt & $^{220}$Hg \\
\underline{$^{35}$P } & $^{46}$Cr & \underline{$^{67}$Cu} & \underline{\textit{$^\textit{84}$Kr}} & \underline{\textit{$^\textit{90}$Zr}} & \underline{$^{106}$Mo} & $^{122}$Cd & $^{132}$Sb & $^{137}$Xe & $^{160}$Nd & $^{160}$Ho & \underline{$^{197}$Lu} & \underline{$^{201}$W } & $^{206}$Os & \underline{$^{212}$Pt} & \underline{$^{222}$Hg} \\
\textit{$^\textit{33}$S } & $^{49}$Cr & $^{68}$Cu & \underline{\textit{$^\textit{86}$Kr}} & \textit{$^\textit{91}$Zr} & $^{108}$Mo & \underline{$^{117}$In} & \underline{$^{133}$Sb} & \underline{$^{138}$Xe} & $^{162}$Nd & $^{162}$Ho & $^{199}$Lu & \underline{$^{202}$W } & $^{207}$Os & $^{214}$Pt & \underline{$^{224}$Hg} \\
\underline{\textit{$^\textit{36}$S }} & $^{51}$Cr & \underline{$^{69}$Cu} & \underline{$^{87}$Kr} & $^{93}$Zr & \underline{$^{102}$Tc} & \underline{$^{119}$In} & $^{135}$Sb & \underline{$^{140}$Xe} & $^{164}$Nd & $^{164}$Ho & $^{188}$Hf & \underline{$^{203}$W } & \underline{$^{208}$Os} & $^{216}$Pt & \underline{$^{226}$Hg} \\
\textit{$^\textit{37}$Cl} & \underline{\textit{$^\textit{52}$Cr}} & $^{71}$Cu & \underline{$^{88}$Kr} & \underline{\textit{$^\textit{94}$Zr}} & $^{104}$Tc & \underline{$^{121}$In} & \underline{\textit{$^\textit{126}$Te}} & \underline{$^{142}$Xe} & $^{131}$Pm & \textit{$^\textit{165}$Ho} & \underline{$^{190}$Hf} & $^{204}$W & \underline{$^{210}$Os} & \underline{$^{218}$Pt} & $^{228}$Hg \\
\underline{$^{39}$Cl} & \textit{$^\textit{53}$Cr} & \underline{\textit{$^\textit{68}$Zn}} & \underline{$^{90}$Kr} & \underline{$^{95}$Zr} & $^{107}$Tc & \underline{$^{123}$In} & \underline{\textit{$^\textit{128}$Te}} & $^{144}$Xe & \underline{$^{153}$Pm} & $^{168}$Ho & \underline{$^{192}$Hf} & $^{205}$W & $^{212}$Os & \underline{$^{220}$Pt} & $^{230}$Hg \\
\underline{$^{39}$Ar} & \underline{\textit{$^\textit{54}$Cr}} & $^{69}$Zn & \underline{$^{84}$Rb} & \underline{\textit{$^\textit{96}$Zr}} & $^{109}$Tc & \underline{$^{125}$In} & \underline{\textit{$^\textit{130}$Te}} & \underline{$^{146}$Xe} & $^{142}$Sm & $^{180}$Er & $^{194}$Hf & \underline{$^{206}$W } & \underline{$^{214}$Os} & \underline{$^{222}$Pt} & $^{232}$Hg \\
\underline{\textit{$^\textit{40}$Ar}} & \underline{$^{56}$Cr} & \underline{\textit{$^\textit{70}$Zn}} & \underline{$^{86}$Rb} & \underline{$^{97}$Zr} & $^{93}$Ru & \underline{$^{127}$In} & \underline{$^{131}$Te} & $^{121}$Cs & $^{162}$Sm & $^{182}$Er & \underline{$^{195}$Hf} & $^{208}$W & \underline{$^{216}$Os} & \underline{$^{224}$Pt} & $^{234}$Hg \\
\textit{$^\textit{39}$K } & $^{47}$Mn & \underline{$^{72}$Zn} & \underline{$^{87}$Rb} & \underline{$^{98}$Zr} & $^{109}$Ru & \textit{$^\textit{118}$Sn} & \underline{$^{132}$Te} & $^{122}$Cs & $^{164}$Sm & \underline{$^{184}$Er} & \underline{$^{196}$Hf} & \underline{$^{210}$W } & $^{218}$Os & $^{226}$Pt & $^{236}$Hg \\
\underline{\textit{$^\textit{41}$K }} & $^{53}$Mn & \underline{$^{74}$Zn} & $^{89}$Rb & \underline{$^{100}$Zr} & \underline{$^{110}$Ru} & \underline{\textit{$^\textit{120}$Sn}} & \underline{$^{133}$Te} & \underline{$^{137}$Cs} & $^{166}$Sm & \underline{$^{186}$Er} & \underline{$^{197}$Hf} & \underline{$^{212}$W } & $^{220}$Os & $^{228}$Pt & $^{238}$Hg \\
\underline{$^{43}$K } & \underline{$^{57}$Mn} & $^{70}$Ga & \underline{\textit{$^\textit{86}$Sr}} & \underline{$^{102}$Zr} & \underline{$^{112}$Ru} & \underline{\textit{$^\textit{122}$Sn}} & \underline{$^{134}$Te} & \underline{\textit{$^\textit{138}$Ba}} & $^{168}$Sm & $^{188}$Er & \underline{$^{198}$Hf} & \underline{$^{214}$W } & $^{186}$Ir & $^{230}$Pt & $^{240}$Hg \\
$^{41}$Ca & $^{53}$Fe & \underline{$^{75}$Ge} & \textit{$^\textit{87}$Sr} & $^{85}$Nb & \underline{$^{111}$Rh} & \underline{\textit{$^\textit{124}$Sn}} & \underline{$^{135}$Te} & $^{139}$Ba & $^{170}$Sm & \underline{$^{194}$Er} & \underline{$^{199}$Hf} & $^{182}$Re & \underline{\textit{$^\textit{193}$Ir}} & \underline{$^{205}$Au} & $^{242}$Hg \\
\textit{$^\textit{42}$Ca} & $^{55}$Fe & \underline{\textit{$^\textit{76}$Ge}} & \underline{\textit{$^\textit{88}$Sr}} & \underline{$^{95}$Nb} & $^{113}$Rh & \underline{$^{126}$Sn} & \underline{$^{136}$Te} & \underline{$^{145}$Ba} & $^{148}$Eu & \underline{$^{196}$Er} & \underline{$^{200}$Hf} & \textit{$^\textit{187}$Re} & \underline{$^{203}$Ir} & \underline{$^{213}$Au} & $^{195}$Tl \\
\underline{\textit{$^\textit{44}$Ca}} & \underline{\textit{$^\textit{58}$Fe}} & \underline{$^{78}$Ge} & $^{89}$Sr & \underline{$^{97}$Nb} & $^{115}$Rh & \underline{$^{128}$Sn} & \underline{$^{137}$Te} & \underline{$^{147}$Ba} & $^{166}$Gd & \underline{$^{195}$Tm} & \underline{$^{201}$Hf} & $^{189}$Re & $^{205}$Ir & \underline{$^{215}$Au} & $^{201}$Tl \\
\underline{\textit{$^\textit{46}$Ca}} & \underline{$^{60}$Fe} & $^{71}$As & \underline{$^{90}$Sr} & \underline{$^{99}$Nb} & \underline{$^{111}$Pd} & \underline{$^{130}$Sn} & \underline{$^{138}$Te} & \underline{$^{148}$Ba} & $^{168}$Gd & $^{197}$Tm & $^{202}$Hf & \underline{$^{201}$Re} & $^{207}$Ir & \underline{\textit{$^\textit{202}$Hg}} & $^{217}$Tl \\
\underline{\textit{$^\textit{48}$Ca}} & $^{53}$Co & \underline{$^{78}$As} & \underline{$^{91}$Sr} & $^{100}$Nb & \underline{$^{112}$Pd} & \underline{$^{131}$Sn} & $^{139}$Te & \underline{$^{150}$Ba} & $^{170}$Gd & $^{160}$Yb & \underline{$^{204}$Hf} & \textit{$^\textit{184}$Os} & \underline{\textit{$^\textit{198}$Pt}} & \underline{$^{206}$Hg} & $^{218}$Tl \\
$^{44}$Sc & $^{55}$Co & \underline{$^{79}$As} & \underline{$^{92}$Sr} & $^{101}$Nb & $^{113}$Pd & \underline{$^{132}$Sn} & $^{140}$Te & \textit{$^\textit{139}$La} & $^{172}$Gd & \underline{$^{186}$Yb} & $^{206}$Hf & \textit{$^\textit{186}$Os} & \underline{$^{201}$Pt} & \underline{$^{207}$Hg} & $^{221}$Tl \\
\underline{$^{47}$Sc} & $^{57}$Co & \underline{$^{81}$As} & \underline{$^{93}$Sr} & \underline{$^{103}$Nb} & \underline{$^{114}$Pd} & \underline{$^{133}$Sn} & $^{142}$Te & $^{124}$Ce & $^{174}$Gd & \underline{$^{188}$Yb} & $^{165}$Ta & \textit{$^\textit{190}$Os} & \underline{$^{202}$Pt} & \underline{$^{208}$Hg} &  \\
\underline{$^{49}$Sc} & \textit{$^\textit{59}$Co} & \underline{$^{79}$Se} & \underline{$^{94}$Sr} & \underline{$^{105}$Nb} & \underline{$^{116}$Pd} & \underline{$^{134}$Sn} & $^{126}$I & $^{141}$Ce & $^{176}$Gd & \underline{$^{190}$Yb} & \underline{$^{199}$Ta} & \underline{$^{195}$Os} & $^{203}$Pt & \underline{$^{209}$Hg} &  \\
\textit{$^\textit{47}$Ti} & \underline{$^{61}$Co} & \underline{\textit{$^\textit{80}$Se}} & \underline{$^{96}$Sr} & $^{91}$Mo & $^{118}$Pd & $^{135}$Sn & \underline{$^{133}$I } & \underline{$^{154}$Ce} & \underline{$^{163}$Tb} & $^{192}$Yb & $^{166}$W & \underline{$^{198}$Os} & \underline{$^{204}$Pt} & \underline{$^{210}$Hg} &  \\
\underline{\textit{$^\textit{48}$Ti}} & \underline{$^{63}$Co} & \underline{\textit{$^\textit{82}$Se}} & $^{79}$Y & $^{93}$Mo & $^{114}$Ag & \underline{$^{136}$Sn} & \underline{$^{135}$I } & $^{156}$Ce & $^{172}$Dy & \underline{$^{194}$Yb} & \textit{$^\textit{184}$W } & \underline{$^{199}$Os} & \underline{$^{205}$Pt} & $^{211}$Hg &
\end{tabular}
\end{ruledtabular}
\end{table*}
\endgroup

\begingroup
\squeezetable
\begin{table*}
\caption{\label{tab:ap} Same as Table \ref{tab:ag} but for ($\alpha$,p).}
\begin{ruledtabular}
\begin{tabular}{cccccccccccccccc}
$^{23}$Ne & $^{55}$Fe & $^{89}$Zn & $^{102}$Se & $^{113}$Sr & $^{111}$Mo & $^{130}$Rh & $^{153}$Cd & \underline{$^{104}$Te} & $^{151}$Xe & $^{177}$Ba & $^{154}$Pr & $^{188}$Sm & $^{200}$Dy & $^{165}$Hf & $^{199}$Pt \\
$^{28}$Na & $^{62}$Fe & $^{91}$Zn & $^{103}$Se & $^{114}$Sr & $^{112}$Mo & $^{132}$Rh & $^{154}$Cd & \underline{$^{106}$Te} & $^{152}$Xe & $^{179}$Ba & $^{156}$Pr & $^{200}$Sm & $^{215}$Dy & $^{167}$Hf & $^{201}$Pt \\
\textit{$^\textit{25}$Mg} & $^{63}$Fe & $^{92}$Zn & $^{104}$Se & $^{115}$Sr & $^{113}$Mo & $^{140}$Rh & $^{155}$Cd & $^{107}$Te & $^{154}$Xe & $^{181}$Ba & $^{194}$Pr & $^{201}$Sm & $^{216}$Dy & \textit{$^\textit{179}$Hf} & $^{203}$Pt \\
$^{37}$Al & $^{69}$Fe & $^{93}$Zn & $^{105}$Se & $^{116}$Sr & $^{114}$Mo & $^{146}$Rh & $^{156}$Cd & $^{109}$Te & $^{155}$Xe & $^{182}$Ba & $^{122}$Nd & $^{207}$Sm & $^{217}$Dy & $^{183}$Hf & $^{205}$Pt \\
\textit{$^\textit{29}$Si} & $^{70}$Fe & $^{94}$Zn & $^{107}$Se & $^{120}$Sr & $^{115}$Mo & $^{148}$Rh & \underline{$^{99}$In} & $^{113}$Te & $^{156}$Xe & $^{184}$Ba & \underline{$^{124}$Nd} & $^{208}$Sm & $^{221}$Dy & $^{185}$Hf & $^{208}$Pt \\
$^{37}$P & $^{71}$Fe & $^{95}$Zn & $^{108}$Se & $^{121}$Sr & $^{117}$Mo & $^{92}$Pd & $^{101}$In & $^{115}$Te & $^{157}$Xe & $^{185}$Ba & $^{129}$Nd & $^{137}$Eu & $^{145}$Ho & $^{186}$Hf & $^{247}$Pt \\
\textit{$^\textit{33}$S } & $^{72}$Fe & $^{96}$Zn & $^{110}$Se & $^{123}$Sr & $^{118}$Mo & $^{94}$Pd & \underline{$^{103}$In} & $^{133}$Te & $^{158}$Xe & $^{186}$Ba & $^{132}$Nd & $^{132}$Gd & $^{147}$Ho & $^{187}$Hf & $^{253}$Pt \\
$^{43}$S & $^{73}$Fe & $^{97}$Zn & $^{111}$Se & $^{124}$Sr & $^{119}$Mo & \underline{$^{95}$Pd} & $^{105}$In & $^{135}$Te & $^{159}$Xe & $^{187}$Ba & $^{133}$Nd & $^{133}$Gd & $^{170}$Ho & $^{188}$Hf & $^{255}$Pt \\
$^{45}$S & $^{74}$Fe & $^{99}$Zn & $^{112}$Se & $^{125}$Sr & $^{120}$Mo & \underline{$^{96}$Pd} & $^{107}$In & $^{136}$Te & $^{160}$Xe & $^{188}$Ba & $^{135}$Nd & \underline{$^{134}$Gd} & $^{172}$Ho & $^{190}$Hf & $^{256}$Pt \\
$^{38}$Cl & $^{79}$Fe & $^{100}$Zn & $^{115}$Se & $^{126}$Sr & $^{121}$Mo & \underline{$^{97}$Pd} & $^{140}$In & $^{137}$Te & $^{161}$Xe & $^{189}$Ba & $^{147}$Nd & $^{135}$Gd & $^{145}$Er & $^{192}$Hf & $^{257}$Pt \\
$^{39}$Cl & $^{82}$Fe & $^{101}$Zn & $^{117}$Se & $^{128}$Sr & $^{123}$Mo & $^{98}$Pd & $^{142}$In & $^{139}$Te & $^{162}$Xe & $^{140}$La & $^{151}$Nd & $^{137}$Gd & $^{146}$Er & $^{193}$Hf & $^{258}$Pt \\
$^{61}$Cl & $^{84}$Fe & $^{103}$Zn & $^{118}$Se & $^{129}$Sr & $^{125}$Mo & $^{101}$Pd & $^{148}$In & $^{140}$Te & $^{164}$Xe & $^{142}$La & $^{153}$Nd & $^{138}$Gd & $^{154}$Er & $^{194}$Hf & $^{259}$Pt \\
$^{37}$Ar & $^{85}$Fe & $^{62}$Ga & $^{92}$Br & $^{110}$Y & $^{127}$Mo & $^{107}$Pd & \underline{$^{99}$Sn} & $^{142}$Te & $^{166}$Xe & $^{144}$La & $^{155}$Nd & $^{140}$Gd & $^{158}$Er & $^{195}$Hf & $^{198}$Au \\
$^{39}$Ar & $^{87}$Fe & \textit{$^\textit{69}$Ga} & $^{108}$Br & $^{126}$Y & $^{128}$Mo & $^{109}$Pd & \underline{$^{100}$Sn} & $^{143}$Te & $^{167}$Xe & $^{148}$La & $^{157}$Nd & $^{142}$Gd & $^{160}$Er & $^{197}$Hf & $^{173}$Hg \\
$^{42}$Ar & $^{88}$Fe & $^{70}$Ga & $^{70}$Kr & \underline{$^{80}$Zr} & $^{130}$Mo & $^{116}$Pd & $^{101}$Sn & $^{144}$Te & $^{168}$Xe & $^{150}$La & $^{158}$Nd & $^{144}$Gd & \textit{$^\textit{162}$Er} & $^{198}$Hf & $^{175}$Hg \\
$^{43}$Ar & $^{89}$Fe & $^{86}$Ga & \underline{$^{72}$Kr} & $^{81}$Zr & $^{132}$Mo & $^{124}$Pd & $^{107}$Sn & $^{145}$Te & $^{169}$Xe & $^{152}$La & $^{160}$Nd & $^{146}$Gd & \textit{$^\textit{164}$Er} & $^{199}$Hf & $^{177}$Hg \\
$^{44}$Ar & \textit{$^\textit{59}$Co} & \underline{$^{62}$Ge} & $^{73}$Kr & \underline{$^{84}$Zr} & $^{133}$Mo & $^{125}$Pd & \textit{$^\textit{117}$Sn} & $^{146}$Te & $^{170}$Xe & $^{154}$La & $^{161}$Nd & $^{151}$Gd & \textit{$^\textit{166}$Er} & $^{231}$Hf & $^{187}$Hg \\
$^{47}$Ar & $^{61}$Co & \underline{$^{64}$Ge} & \underline{$^{74}$Kr} & $^{85}$Zr & $^{134}$Mo & $^{127}$Pd & \textit{$^\textit{118}$Sn} & $^{147}$Te & $^{171}$Xe & $^{190}$La & $^{162}$Nd & $^{161}$Gd & \textit{$^\textit{168}$Er} & $^{233}$Hf & \textit{$^\textit{199}$Hg} \\
$^{57}$Ar & $^{63}$Co & \underline{$^{66}$Ge} & $^{75}$Kr & \underline{$^{86}$Zr} & $^{135}$Mo & $^{129}$Pd & $^{121}$Sn & $^{148}$Te & $^{172}$Xe & \underline{$^{116}$Ce} & $^{163}$Nd & $^{162}$Gd & \textit{$^\textit{170}$Er} & $^{235}$Hf & $^{209}$Hg \\
\textit{$^\textit{39}$K } & $^{71}$Co & $^{68}$Ge & \textit{$^\textit{80}$Kr} & \underline{$^{87}$Zr} & $^{136}$Mo & $^{131}$Pd & $^{125}$Sn & $^{149}$Te & $^{173}$Xe & \underline{$^{118}$Ce} & $^{164}$Nd & $^{164}$Gd & $^{171}$Er & $^{158}$Ta & $^{214}$Hg \\
\underline{$^{39}$Ca} & \underline{$^{56}$Ni} & \textit{$^\textit{70}$Ge} & \textit{$^\textit{82}$Kr} & $^{88}$Zr & $^{137}$Mo & $^{132}$Pd & $^{131}$Sn & $^{150}$Te & $^{174}$Xe & \underline{$^{120}$Ce} & $^{165}$Nd & $^{165}$Gd & $^{172}$Er & $^{157}$W & $^{243}$Hg \\
\textit{$^\textit{40}$Ca} & \underline{$^{57}$Ni} & $^{82}$Ge & $^{85}$Kr & $^{89}$Zr & $^{143}$Mo & $^{134}$Pd & $^{132}$Sn & $^{152}$Te & $^{175}$Xe & $^{124}$Ce & $^{166}$Nd & $^{166}$Gd & $^{173}$Er & $^{158}$W & $^{246}$Hg \\
\underline{$^{41}$Ca} & \textit{$^\textit{58}$Ni} & $^{85}$Ge & $^{91}$Kr & \textit{$^\textit{90}$Zr} & $^{144}$Mo & $^{135}$Pd & $^{133}$Sn & $^{153}$Te & $^{176}$Xe & $^{125}$Ce & $^{167}$Nd & $^{167}$Gd & $^{174}$Er & $^{161}$W & $^{247}$Hg \\
\textit{$^\textit{42}$Ca} & $^{59}$Ni & $^{87}$Ge & $^{92}$Kr & \textit{$^\textit{91}$Zr} & $^{128}$Tc & $^{136}$Pd & $^{134}$Sn & $^{154}$Te & $^{179}$Xe & $^{127}$Ce & $^{168}$Nd & $^{170}$Gd & $^{175}$Er & $^{166}$W & $^{248}$Hg \\
\textit{$^\textit{43}$Ca} & $^{71}$Ni & $^{88}$Ge & $^{93}$Kr & $^{99}$Zr & $^{134}$Tc & $^{137}$Pd & $^{135}$Sn & $^{155}$Te & $^{180}$Xe & $^{131}$Ce & $^{170}$Nd & $^{171}$Gd & $^{176}$Er & $^{171}$W & $^{249}$Hg \\
\textit{$^\textit{44}$Ca} & $^{72}$Ni & $^{89}$Ge & $^{94}$Kr & $^{103}$Zr & $^{88}$Ru & $^{138}$Pd & $^{136}$Sn & $^{156}$Te & $^{181}$Xe & $^{139}$Ce & $^{172}$Nd & $^{172}$Gd & $^{178}$Er & $^{190}$W & $^{253}$Hg \\
$^{45}$Ca & $^{73}$Ni & $^{91}$Ge & $^{95}$Kr & $^{104}$Zr & $^{89}$Ru & $^{139}$Pd & $^{137}$Sn & $^{157}$Te & \underline{$^{111}$Cs} & $^{144}$Ce & $^{173}$Nd & $^{176}$Gd & $^{180}$Er & $^{191}$W & $^{255}$Hg \\
$^{47}$Ca & $^{74}$Ni & $^{92}$Ge & $^{96}$Kr & $^{105}$Zr & $^{90}$Ru & $^{140}$Pd & $^{138}$Sn & $^{158}$Te & $^{119}$Cs & $^{148}$Ce & $^{175}$Nd & $^{177}$Gd & $^{181}$Er & $^{192}$W & $^{256}$Hg \\
$^{49}$Ca & $^{75}$Ni & $^{93}$Ge & $^{97}$Kr & $^{106}$Zr & $^{91}$Ru & $^{141}$Pd & $^{139}$Sn & $^{159}$Te & $^{156}$Cs & $^{150}$Ce & $^{176}$Nd & $^{178}$Gd & $^{182}$Er & $^{194}$W & $^{257}$Hg \\
$^{64}$Ca & $^{77}$Ni & $^{95}$Ge & $^{99}$Kr & $^{107}$Zr & \underline{$^{92}$Ru} & $^{142}$Pd & $^{140}$Sn & $^{160}$Te & $^{112}$Ba & $^{151}$Ce & $^{177}$Nd & $^{179}$Gd & $^{183}$Er & $^{195}$W & $^{259}$Hg \\
$^{65}$Ca & $^{78}$Ni & $^{97}$Ge & $^{100}$Kr & $^{109}$Zr & \underline{$^{93}$Ru} & $^{143}$Pd & $^{141}$Sn & $^{162}$Te & $^{114}$Ba & $^{153}$Ce & $^{178}$Nd & $^{181}$Gd & $^{184}$Er & $^{196}$W & $^{261}$Hg \\
$^{66}$Ca & $^{84}$Ni & $^{99}$Ge & $^{101}$Kr & $^{110}$Zr & \underline{$^{94}$Ru} & $^{144}$Pd & $^{142}$Sn & $^{163}$Te & $^{120}$Ba & $^{154}$Ce & $^{180}$Nd & $^{182}$Gd & $^{185}$Er & $^{197}$W & $^{263}$Hg \\
$^{67}$Ca & $^{85}$Ni & $^{100}$Ge & $^{102}$Kr & $^{112}$Zr & $^{95}$Ru & $^{146}$Pd & $^{144}$Sn & $^{164}$Te & $^{121}$Ba & $^{155}$Ce & $^{181}$Nd & $^{184}$Gd & $^{188}$Er & $^{199}$W & $^{265}$Hg \\
$^{68}$Ca & $^{86}$Ni & $^{101}$Ge & $^{103}$Kr & $^{113}$Zr & \textit{$^\textit{96}$Ru} & $^{147}$Pd & $^{145}$Sn & $^{166}$Te & $^{122}$Ba & $^{156}$Ce & $^{182}$Nd & $^{186}$Gd & $^{190}$Er & $^{200}$W & $^{202}$Tl \\
$^{69}$Ca & $^{87}$Ni & $^{103}$Ge & $^{104}$Kr & $^{114}$Zr & \textit{$^\textit{99}$Ru} & $^{148}$Pd & $^{146}$Sn & $^{167}$Te & $^{123}$Ba & $^{158}$Ce & $^{185}$Nd & $^{187}$Gd & $^{191}$Er & $^{201}$W & $^{179}$Pb \\
$^{49}$Sc & $^{88}$Ni & $^{104}$Ge & $^{105}$Kr & $^{115}$Zr & $^{107}$Ru & $^{149}$Pd & $^{147}$Sn & $^{168}$Te & \textit{$^\textit{134}$Ba} & $^{159}$Ce & $^{186}$Nd & $^{189}$Gd & $^{193}$Er & $^{202}$W & $^{180}$Pb \\
$^{44}$Ti & $^{89}$Ni & $^{105}$Ge & $^{108}$Kr & $^{116}$Zr & $^{109}$Ru & $^{150}$Pd & $^{148}$Sn & $^{169}$Te & \textit{$^\textit{137}$Ba} & $^{161}$Ce & $^{187}$Nd & $^{190}$Gd & $^{217}$Er & $^{232}$W & $^{181}$Pb \\
\textit{$^\textit{46}$Ti} & $^{91}$Ni & $^{106}$Ge & $^{109}$Kr & $^{117}$Zr & $^{114}$Ru & $^{151}$Pd & $^{150}$Sn & $^{170}$Te & $^{141}$Ba & $^{162}$Ce & $^{188}$Nd & $^{192}$Gd & $^{218}$Er & $^{241}$W & $^{182}$Pb \\
\textit{$^\textit{48}$Ti} & $^{92}$Ni & $^{108}$Ge & $^{113}$Kr & $^{118}$Zr & $^{115}$Ru & $^{152}$Pd & $^{151}$Sn & $^{171}$Te & $^{146}$Ba & $^{163}$Ce & $^{194}$Nd & $^{198}$Gd & $^{219}$Er & $^{160}$Re & $^{185}$Pb \\
\textit{$^\textit{49}$Ti} & $^{93}$Ni & $^{109}$Ge & $^{114}$Kr & $^{121}$Zr & $^{117}$Ru & $^{153}$Pd & $^{152}$Sn & $^{173}$Te & $^{147}$Ba & $^{164}$Ce & $^{196}$Nd & $^{202}$Gd & $^{226}$Er & $^{184}$Re & $^{187}$Pb \\
\textit{$^\textit{50}$Ti} & $^{95}$Ni & $^{111}$Ge & $^{115}$Kr & $^{122}$Zr & $^{119}$Ru & \underline{$^{97}$Ag} & $^{153}$Sn & $^{176}$Te & $^{148}$Ba & $^{165}$Ce & $^{131}$Pm & $^{208}$Gd & $^{227}$Er & $^{194}$Re & $^{189}$Pb \\
$^{51}$Ti & $^{97}$Ni & $^{112}$Ge & $^{117}$Kr & $^{123}$Zr & $^{121}$Ru & $^{99}$Ag & $^{154}$Sn & $^{136}$I & $^{149}$Ba & $^{166}$Ce & $^{158}$Pm & $^{209}$Gd & \underline{$^{145}$Tm} & $^{165}$Os & $^{191}$Pb \\
$^{62}$Ti & $^{99}$Ni & $^{68}$As & $^{119}$Kr & $^{125}$Zr & $^{125}$Ru & $^{123}$Ag & $^{155}$Sn & $^{140}$I & $^{150}$Ba & $^{168}$Ce & $^{160}$Pm & $^{211}$Gd & \underline{$^{147}$Tm} & $^{168}$Os & $^{195}$Pb \\
$^{63}$Ti & \underline{$^{58}$Cu} & $^{74}$As & $^{123}$Kr & $^{126}$Zr & $^{126}$Ru & $^{125}$Ag & $^{156}$Sn & $^{142}$I & $^{151}$Ba & $^{170}$Ce & $^{128}$Sm & $^{213}$Gd & $^{151}$Yb & $^{170}$Os & $^{197}$Pb \\
$^{64}$Ti & $^{59}$Cu & $^{78}$As & $^{124}$Kr & $^{127}$Zr & $^{127}$Ru & $^{140}$Ag & $^{157}$Sn & $^{154}$I & $^{152}$Ba & $^{171}$Ce & \underline{$^{130}$Sm} & $^{166}$Tb & $^{153}$Yb & $^{173}$Os & $^{199}$Pb \\
$^{66}$Ti & $^{61}$Cu & $^{90}$As & $^{80}$Rb & $^{128}$Zr & $^{129}$Ru & $^{97}$Cd & $^{158}$Sn & $^{168}$I & $^{153}$Ba & $^{172}$Ce & $^{131}$Sm & $^{168}$Tb & $^{154}$Yb & $^{180}$Os & $^{201}$Pb \\
$^{68}$Ti & \textit{$^\textit{63}$Cu} & $^{66}$Se & $^{124}$Rb & $^{129}$Zr & $^{130}$Ru & $^{130}$Cd & $^{160}$Sn & $^{172}$I & $^{154}$Ba & $^{173}$Ce & $^{132}$Sm & $^{137}$Dy & $^{155}$Yb & $^{182}$Os & $^{213}$Pb \\
$^{69}$Ti & \textit{$^\textit{65}$Cu} & \underline{$^{68}$Se} & \underline{$^{80}$Sr} & $^{130}$Zr & $^{132}$Ru & $^{131}$Cd & $^{161}$Sn & $^{110}$Xe & $^{156}$Ba & $^{174}$Ce & $^{133}$Sm & $^{139}$Dy & $^{157}$Yb & $^{195}$Os & $^{214}$Pb \\
\textit{$^\textit{51}$V } & $^{67}$Cu & \underline{$^{70}$Se} & $^{82}$Sr & $^{131}$Zr & $^{133}$Ru & $^{133}$Cd & $^{162}$Sn & \underline{$^{112}$Xe} & $^{157}$Ba & $^{175}$Ce & $^{157}$Sm & $^{140}$Dy & $^{160}$Yb & $^{196}$Os & $^{216}$Pb \\
\textit{$^\textit{50}$Cr} & $^{69}$Cu & $^{72}$Se & \textit{$^\textit{84}$Sr} & $^{132}$Zr & $^{134}$Ru & $^{135}$Cd & $^{163}$Sn & $^{115}$Xe & $^{158}$Ba & $^{176}$Ce & $^{159}$Sm & $^{141}$Dy & $^{161}$Yb & $^{198}$Os & $^{246}$Pb \\
$^{51}$Cr & $^{71}$Cu & \textit{$^\textit{74}$Se} & $^{85}$Sr & $^{133}$Zr & $^{135}$Ru & $^{136}$Cd & $^{165}$Sn & \textit{$^\textit{126}$Xe} & $^{159}$Ba & $^{177}$Ce & $^{160}$Sm & $^{143}$Dy & $^{185}$Yb & $^{201}$Os & $^{252}$Pb \\
\textit{$^\textit{52}$Cr} & \underline{$^{58}$Zn} & $^{85}$Se & \textit{$^\textit{86}$Sr} & $^{126}$Nb & $^{136}$Ru & $^{137}$Cd & $^{167}$Sn & \textit{$^\textit{131}$Xe} & $^{161}$Ba & $^{178}$Ce & $^{163}$Sm & $^{145}$Dy & $^{187}$Yb & $^{203}$Os & $^{253}$Pb \\
\textit{$^\textit{53}$Cr} & \underline{$^{59}$Zn} & $^{87}$Se & \textit{$^\textit{87}$Sr} & $^{128}$Nb & $^{137}$Ru & $^{138}$Cd & $^{168}$Sn & $^{133}$Xe & $^{162}$Ba & $^{179}$Ce & $^{164}$Sm & $^{146}$Dy & $^{188}$Yb & $^{206}$Os & $^{254}$Pb \\
$^{65}$Cr & \underline{$^{60}$Zn} & $^{88}$Se & \textit{$^\textit{88}$Sr} & $^{85}$Mo & $^{138}$Ru & $^{140}$Cd & $^{169}$Sn & $^{135}$Xe & $^{163}$Ba & $^{183}$Ce & $^{165}$Sm & $^{151}$Dy & $^{190}$Yb & $^{250}$Os & $^{257}$Pb \\
$^{66}$Cr & \underline{$^{62}$Zn} & $^{89}$Se & $^{89}$Sr & \underline{$^{86}$Mo} & $^{139}$Ru & $^{141}$Cd & \underline{$^{105}$Sb} & $^{137}$Xe & $^{164}$Ba & $^{184}$Ce & $^{166}$Sm & $^{169}$Dy & $^{192}$Yb & $^{251}$Os & $^{258}$Pb \\
$^{67}$Cr & \textit{$^\textit{64}$Zn} & $^{90}$Se & $^{102}$Sr & \underline{$^{88}$Mo} & $^{140}$Ru & $^{142}$Cd & $^{106}$Sb & $^{139}$Xe & $^{166}$Ba & $^{185}$Ce & $^{169}$Sm & $^{171}$Dy & $^{193}$Yb & $^{253}$Os & $^{259}$Pb \\
$^{68}$Cr & $^{72}$Zn & $^{91}$Se & $^{103}$Sr & \underline{$^{89}$Mo} & $^{141}$Ru & $^{143}$Cd & \underline{$^{107}$Sb} & $^{141}$Xe & $^{168}$Ba & $^{186}$Ce & $^{172}$Sm & $^{173}$Dy & $^{195}$Yb & $^{184}$Ir & $^{260}$Pb \\
$^{69}$Cr & $^{77}$Zn & $^{92}$Se & $^{104}$Sr & $^{90}$Mo & $^{143}$Ru & $^{144}$Cd & $^{111}$Sb & $^{142}$Xe & $^{169}$Ba & $^{187}$Ce & $^{174}$Sm & $^{179}$Dy & $^{229}$Yb & $^{186}$Ir & $^{261}$Pb \\
$^{70}$Cr & $^{79}$Zn & $^{93}$Se & $^{105}$Sr & \underline{$^{91}$Mo} & $^{144}$Ru & $^{145}$Cd & $^{113}$Sb & $^{143}$Xe & $^{170}$Ba & $^{188}$Ce & $^{175}$Sm & $^{181}$Dy & $^{231}$Yb & $^{174}$Pt & $^{262}$Pb \\
$^{71}$Cr & $^{80}$Zn & $^{94}$Se & $^{107}$Sr & \textit{$^\textit{92}$Mo} & $^{145}$Ru & $^{146}$Cd & $^{115}$Sb & $^{144}$Xe & $^{171}$Ba & $^{190}$Ce & $^{177}$Sm & $^{182}$Dy & $^{182}$Lu & $^{176}$Pt & $^{263}$Pb \\
$^{77}$Cr & $^{84}$Zn & $^{95}$Se & $^{108}$Sr & $^{93}$Mo & $^{146}$Ru & $^{148}$Cd & $^{138}$Sb & $^{146}$Xe & $^{172}$Ba & $^{191}$Ce & $^{178}$Sm & $^{186}$Dy & $^{184}$Lu & $^{178}$Pt &  \\
$^{79}$Cr & $^{85}$Zn & $^{97}$Se & $^{109}$Sr & \textit{$^\textit{98}$Mo} & $^{147}$Ru & $^{149}$Cd & $^{140}$Sb & $^{147}$Xe & $^{173}$Ba & $^{192}$Ce & $^{180}$Sm & $^{188}$Dy & $^{154}$Hf & $^{180}$Pt &  \\
$^{82}$Cr & $^{86}$Zn & $^{98}$Se & $^{110}$Sr & $^{103}$Mo & $^{148}$Ru & $^{150}$Cd & $^{146}$Sb & $^{148}$Xe & $^{174}$Ba & $^{193}$Ce & $^{182}$Sm & $^{189}$Dy & $^{157}$Hf & $^{182}$Pt &  \\
$^{54}$Mn & $^{87}$Zn & $^{99}$Se & $^{111}$Sr & $^{109}$Mo & $^{149}$Ru & $^{151}$Cd & $^{168}$Sb & $^{149}$Xe & $^{175}$Ba & $^{129}$Pr & $^{184}$Sm & $^{191}$Dy & $^{158}$Hf & $^{184}$Pt &  \\
\textit{$^\textit{54}$Fe} & $^{88}$Zn & $^{101}$Se & $^{112}$Sr & $^{110}$Mo & $^{95}$Rh & $^{152}$Cd & $^{170}$Sb & $^{150}$Xe & $^{176}$Ba & $^{152}$Pr & $^{185}$Sm & $^{192}$Dy & $^{159}$Hf & $^{186}$Pt &
\end{tabular}
\end{ruledtabular}
\end{table*}
\endgroup


\begingroup
\squeezetable
\begin{table}
\caption{\label{tab:pa} Same as Table \ref{tab:ag} but for (p,$\alpha$).}
\begin{ruledtabular}
\begin{tabular}{cccccccc}
$^{31}$Si & $^{48}$Ar & $^{62}$Ar & $^{70}$Ca & $^{70}$Se & $^{86}$Zr & \textit{$^\textit{92}$Mo} &  \\
$^{34}$Si & $^{56}$Ar & $^{63}$Ar & $^{51}$Ti & $^{116}$Se & $^{87}$Zr & $^{86}$Tc &  \\
$^{46}$Si & $^{58}$Ar & \textit{$^\textit{43}$Ca} & $^{72}$Ti & $^{70}$Br & $^{80}$Mo & $^{91}$Ru &  \\
$^{46}$S & $^{59}$Ar & $^{47}$Ca & $^{48}$Mn & $^{74}$Kr & $^{89}$Mo & $^{93}$Ru &  \\
$^{44}$Ar & $^{60}$Ar & $^{68}$Ca & $^{57}$Zn & $^{74}$Rb & $^{90}$Mo & $^{95}$Cd &  \\
$^{47}$Ar & $^{61}$Ar & $^{69}$Ca & $^{61}$Ge & \textit{$^\textit{86}$Sr} & $^{91}$Mo & $^{99}$Cd &
\end{tabular}
\end{ruledtabular}
\end{table}
\endgroup

\section{Summary}

A previous investigation was improved by studying the g.s.\ contributions in forward and reverse reaction directions instead of comparing the SEFs. Again, it was found that -- contrary to common wisdom -- some endothermic reactions exhibit smaller g.s.\ contributions to the stellar reaction rate than their exothermic counterparts and are thus preferable for experimental and theoretical studies, if one of the reaction directions is found to be important in a hot plasma. The main cause of suppression of the excited state contributions in an endothermic reaction is the Coulomb suppression of transitions with low relative interaction energy \cite{coulsupp}. The larger number of cases is due to the different temperature dependence of the g.s.\ contributions compared to the SEFs.

For experimental application, it is advisable to first search for a reaction with astrophysical importance and then to check in the above tables whether it belongs to the exceptions for which the experimentally accessible g.s.\ contribution to the stellar rate is larger in the endothermic direction. Otherwise, a determination of the cross sections of the exothermic reaction would be preferrable for astrophysical application, if feasible.

Suppression of thermally excited state contributions is not limited to reactions treated in the Hauser-Feshbach model but also appears in resonant reactions and in direct reactions and similar considerations apply as those discussed above.

\acknowledgments
This work was supported by the Hungarian Academy of Sciences, the ESF EUROCORES programme EuroGENESIS, the THEXO collaboration within the 7th Framework Programme of the EU, the European Research Council, and the Swiss NSF.

\end{document}